\documentclass[twocolumn,trackchanges,lineno]{aastex6}
\usepackage{amsmath}
\usepackage{amssymb}
\usepackage{amsthm}
\usepackage{natbib,aasdefs,url,bm}
\usepackage{array}
\usepackage{float}
\usepackage{graphicx}
\usepackage{subfigure}
\usepackage{color}

\newcommand{\msun}{\rm M_{\odot}}

\newcommand{\mdotenv}{\dot M_{\rm env}}
\newcommand{\mdotw}{\dot M_w}

\begin{document}

\title{Outflow-Driven Transients from the Birth of Binary Black Holes I:\\ Tidally-Locked Secondary Supernovae}

\author{Shigeo S. Kimura\altaffilmark{1,2,3}, Kohta Murase\altaffilmark{1,2,3,4}, and Peter M\'{e}sz\'{a}ros\altaffilmark{1,2,3}}
%\author{Shigeo S. Kimura\altaffilmark{1,2,3}, Kohta Murase\altaffilmark{1,2,3,4}, Peter M\'{e}sz\'{a}ros\altaffilmark{1,2,3}, and Kunihito Ioka\altaffilmark{4}}
\altaffiltext{1}{Department of Physics, Pennsylvania State University, University Park, Pennsylvania 16802, USA}
\altaffiltext{2}{Department of Astronomy \& Astrophysics, Pennsylvania State UNiversity, University Park, Pennsylvania 16802, USA}
\altaffiltext{3}{Center for Particle and Gravitational Astrophysics, Pennsylvania State University, University Park, Pennsylvania 16802, USA}
\altaffiltext{4}{Yukawa Institute for Theoretical Physics, Kyoto, Kyoto 606-8502, Japan}
%\altaffiltext{1}{szk323@psu.edu}
%\altaffiltext{2}{Yukawa Institute for Theoretical Physics, Kyoto, Kyoto 606-8502 Japan}
%% 
%
\begin{abstract}
We propose a new type of electromagnetic transients associated with the birth of binary black holes (BBHs), which may lead to merger events accompanied by gravitational waves in $\sim0.1-1$~Gyr. We consider the newborn BBHs formed through the evolution of isolated massive stellar binaries. For  a close massive binary, consisting of a primary black hole (BH) and a secondary Wolf-Rayet (WR) star that are orbiting around each other, the spin period of the WR star can be tidally synchronized to its orbital period. Then, the angular momentum of the outer material of the WR star is large enough to form an accretion disk around a newborn, secondary BH, following its core-collapse. This disk may produce an energetic outflow with a kinetic energy of $\sim10^{50}-10^{52}$~erg and an outflow velocity of $\sim10^{10}\rm~cm~s^{-1}$, resulting in an optical transient with an absolute magnitude from $\sim -14$ to $\sim-17$ with a duration of around a day. This type of transient also produces detectable radio signals $\sim1-10$ years after the birth of BBHs, via synchrotron emission from non-thermal electrons accelerated at external shocks. 
The predicted optical transients have a shorter duration than ordinary core-collapse supernovae. 
Dedicated optical transient surveys can detect them, and distinguish it from ordinary SNe using the different features of its light curve and late-time spectrum.
In the paper I, we investigate disk-driven outflows from the secondary BH, whereas possible signals from the primary BH will be discussed in the paper II. 
\end{abstract}


\section{Introduction} \label{sec:intro}

The advanced Laser Interferometer Gravitational-wave Observatory (LIGO) revealed the existence of black holes (BH) of $\sim30\rm\msun$ through the detection of gravitational waves (GWs) from mergers of binary black holes (BBHs) \citep{LIGO16e,LIGO16d,LIGO16a}.
The origin of BBHs is under active debate, and several scenarios have been proposed,
such as primordial black hole binaries \citep[e.g.,][]{NST97a,SST16a,MBC16a}, multi-body interactions in star clusters \citep[e.g.,][]{SH93a,PM00a,RHC16a}, and evolution of field binaries \citep[e.g.,][]{TY93a,KIH14a,BHB16a,MLP16a,MM16a}.

Future GW observations may provide the mass, spin, and redshift distributions of merging BBHs,
which are useful to probe the environments where BBHs are formed~ \citep[e.g.,][]{KZK16a,HP17a,FSM17a,ZKK17a}.
Searching for electromagnetic (EM) counterparts from merging BBHs is another way to study them.
However, there is a substantial time gap, typically $\sim0.1-10$ Gyr, between the merger events and the formation of BBHs,
which makes it difficult to probe the environments of the BBH formation.
Besides, simultaneous detections of EM counterparts and GWs are not guaranteed,
because the possible EM signals considered so far require some specific conditions, such as the existence of a fossil disk \citep{PLG16a,MKM16a,KTT17a,IMT17a}, or a BBH formation in anomalously dense environments, such as the inside of very massive stars \citep{Loe16a,DMM17a} or active galactic nuclei \citep{BKH17a,SMH17a}. Instead, searching for the EM radiation from newborn BBHs would enable us more directly to probe the environment of BBH formation.

In this work, we suggest a new class of transient associated with the birth of a BBH system,
which is a natural consequence of the evolution of the progenitor binary system from massive stars.
The transients investigated in this work are {\it not coincident} with the GW emission at the BH-BH merger.
Nevertheless, successful observations can provide important clues about the formation scenario of BBHs.
We describe the basic binary evolution process and the possible outcomes in Section \ref{sec:scenario}, where we consider two scenarios; one powered by the secondary BH and the other powered by the primary BH.
In this paper I, we focus on the transient events driven by the secondary BH. 
We analytically estimate observable features of optical transients in Section \ref{sec:optical}, and the associated radio transients in Section \ref{sec:radio}.  
In Section \ref{sec:summary}, we summarize and discuss our results, including the observational prospects. 
The other type of transients powered by the primary BH are discussed in Paper II \citep{KMM17a}.
We use the notation of $A=A_x10^x$ throughout this work.

\begin{figure*}[tbp]
\begin{center}
\includegraphics[width=\linewidth]{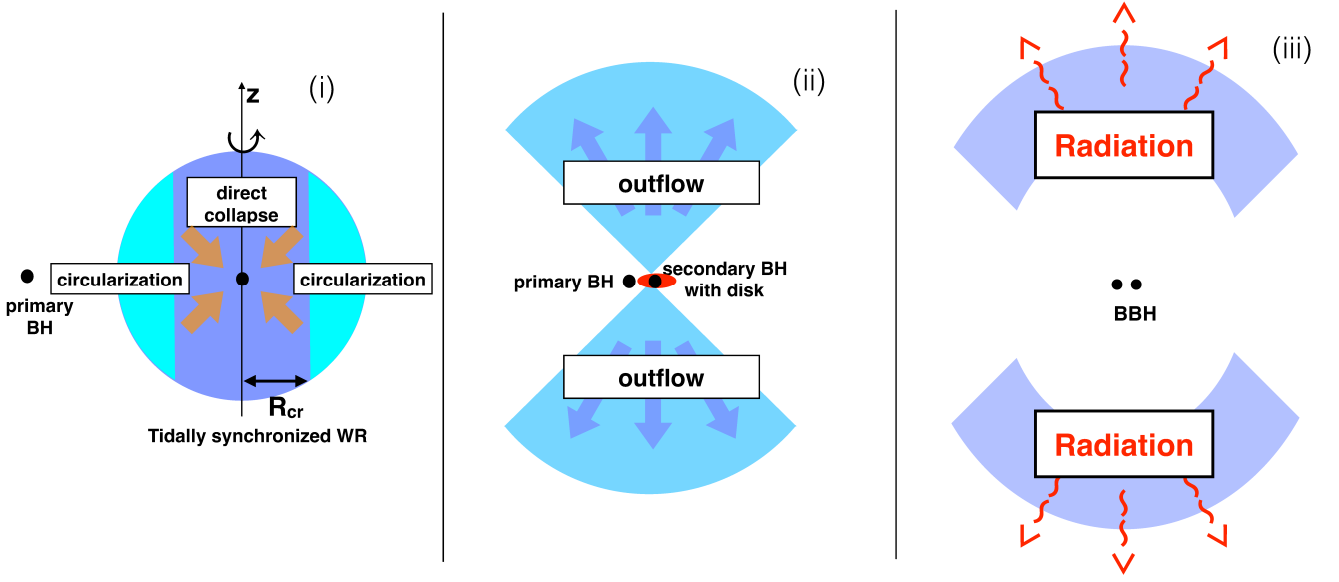}
\caption{Schematic picture of tidally-locked secondary supernovae (TLSSNe); (i) A WR is synchronized via the tidal force before its collapse. The inner part forms a secondary BH, while the outer material forms a disk around the BH. (ii) An ejecta is launched by the disk-driven outflow. (iii) Thermal photons diffuse out from the ejecta.}
    \label{fig:schematic}
   \end{center}
  \end{figure*}

\section{Consequences of the Evolutionary Scenario}\label{sec:scenario}

According to the isolated binary evolution models \citep[e.g.,][]{BHB16a}, the heavier primary collapses to a BH earlier than the secondary, which forms a BH and main-sequence star binary. 
When the secondary evolves to a giant star, the binary separation decreases considerably during a common envelope evolution \citep{Pac76a,Web84a}. 
After that, the secondary is expected to become a Wolf-Rayet star (WR), following the envelope ejection, and the BH-WR binary becomes a BBH after the gravitational collapse of the WR \citep[e.g.][]{DBF12a}. 

At the end of the binary evolution, after the massive secondary star has collapsed and the BBH has formed, its subsequent fate depends on the angular momentum of the secondary star. 
Although the angular momentum distribution of the secondary is highly uncertain, the spin of the secondary star may be tidally locked in a close binary system \citep{Zah77a,Tas87a}. \citet{KZK16a} give the synchronization time as
\begin{equation}
t_{\rm TL}\sim 7.5\times10^4 \left(\frac{t_{\rm mer}}{10^8\rm~yr}\right)^{17/8}\rm~yr,%q^{-1/8}\left(\frac{1+q}{2q}\right)^{31/24} \rm~yr,
\end{equation}
where $t_{\rm mer}=5c^5a^4/(512G^3M_*^3)\sim 1.0\times10^8a_{12}^4M_{*,1.5}^{-3}$ yr is the GW inspiral time, $M_*$ is the primary mass, and $a$ is the binary separation. 
We assume the mass ratio $q=1$ for simplicity and use $M_*\sim10^{1.5}~\msun$ and $a\sim10^{12}$ cm for the purpose of an estimate, 
which indicates that $t_{\rm TL}$ is shorter than the typical lifetime of massive stars, $t_{\rm life}\sim 10^6$ yr. 
However, for a low mass $M_*\sim 10~\msun$ or large separation $a\sim 3\times10^{12}$ cm, $t_{\rm TL}>t_{\rm life}$ is possible. 
We caution that this timescale has significant uncertainties caused by the strong dependence on the detailed stellar structure, 
especially the size of the convective region \citep{KZK16a}. 
The size of the star also affects this timescale.

When $t_{\rm TL}<t_{\rm life}$, the spin period of the secondary is synchronized to its orbital period. 
The spin angular momentum of the WR is high enough to prevent the WR from directly collapsing to a BH. 
The outer region of the WR forms an accretion disk around the newborn secondary BH. 
The accretion rate is high enough to produce radiation-driven powerful outflows  \citep[e.g.][]{OMN05a,JSD14a}, leading to a tidally locked secondary supernova (TLSSN, see Figure \ref{fig:schematic} for the schematic picture). 
The kinetic energy of this outflow is so large that we can expect a radio afterglow. 

In the opposite case when $t_{\rm TL}>t_{\rm life}$, the WR is likely to spin slowly enough to collapse to a BH directly
\footnote{The centrifugal and Colioris forces do not affect a disk formation process when the WR star collapses even when the binary separation is considerably close.}.
When the WR collapses, the outer envelope of the WR is ejected due to energy losses by neutrinos \citep{Nad80a}. 
The primary BH accretes the ejected material, and may produce powerful outflows owing to its high accretion rate. 
This outflow energizes the ejecta and could lead to a primary-induced accretion transient (PIATs), which is discussed in the accompanying paper (Paper II).

A disk-driven outflow can produce a super-luminous supernova \citep{DK13a}, a hypernova \citep{MW99a}, and an optical transient during a single BH formation~\citep{KQ15a} and BH mergers~\citep{MKM16a}. 
TLLSNe and PIATs provide different examples associated with a newborn BBH. 

\section{Optical Emission from TLSSNe}\label{sec:optical}

  \begin{figure}[tbp]
   \begin{center}
   \includegraphics[width=\linewidth]{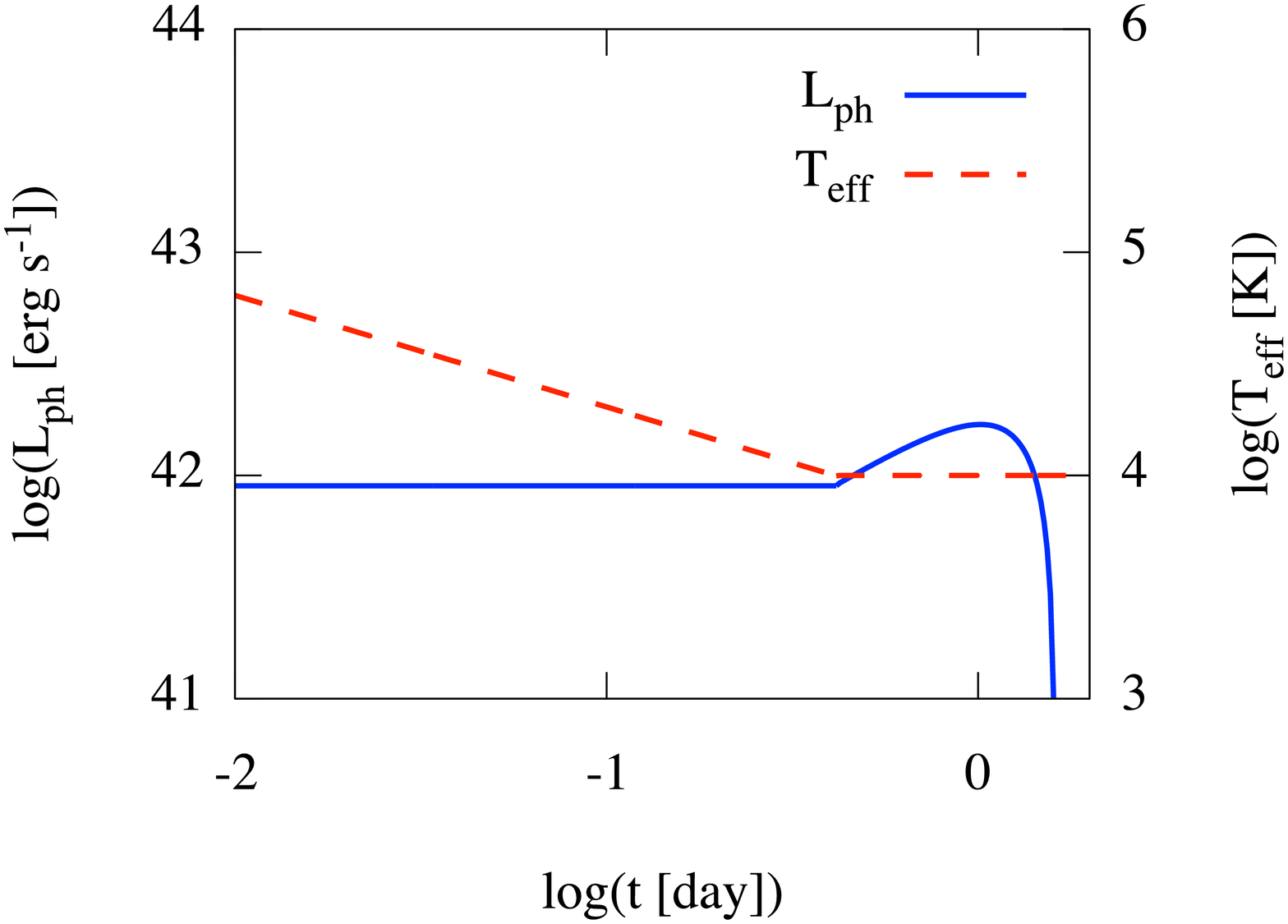}
   \includegraphics[width=\linewidth]{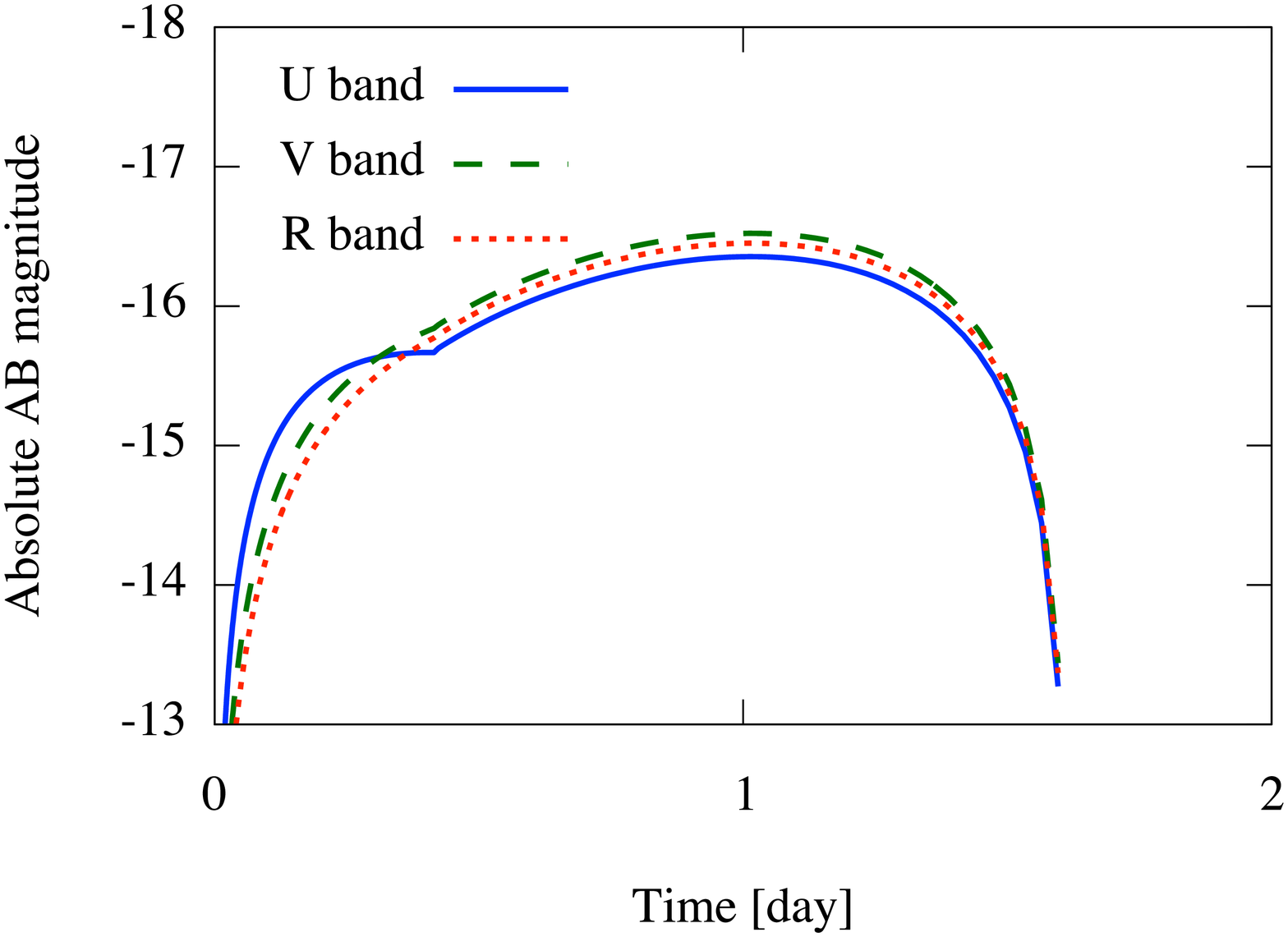}
    \caption{Time evolution of physical quantities for a TLSSN with the fiducial parameter set ($M_*=30\rm~M_{\odot}$, $R_*=10^{11}$ cm, $a=10^{12}$ cm, $\eta_w=10^{-0.5}$, $V_w=10^{10}\rm~cm~s^{-1}$). 
The upper panel indicates the bolometric luminosity of diffusing photons $L_{\rm ph}$ (blue-solid) and the effective temperature $T_{\rm eff}$ (red-dashed). The lower panel shows absolute AB magnitudes for $U$ (blue-solid), $V$ (green-dashed), and $R$ band (red-dotted).}
    \label{fig:tidal_evolv}
   \end{center}
  \end{figure}

  \begin{figure}[tbp]
   \begin{center}
   \includegraphics[width=\linewidth]{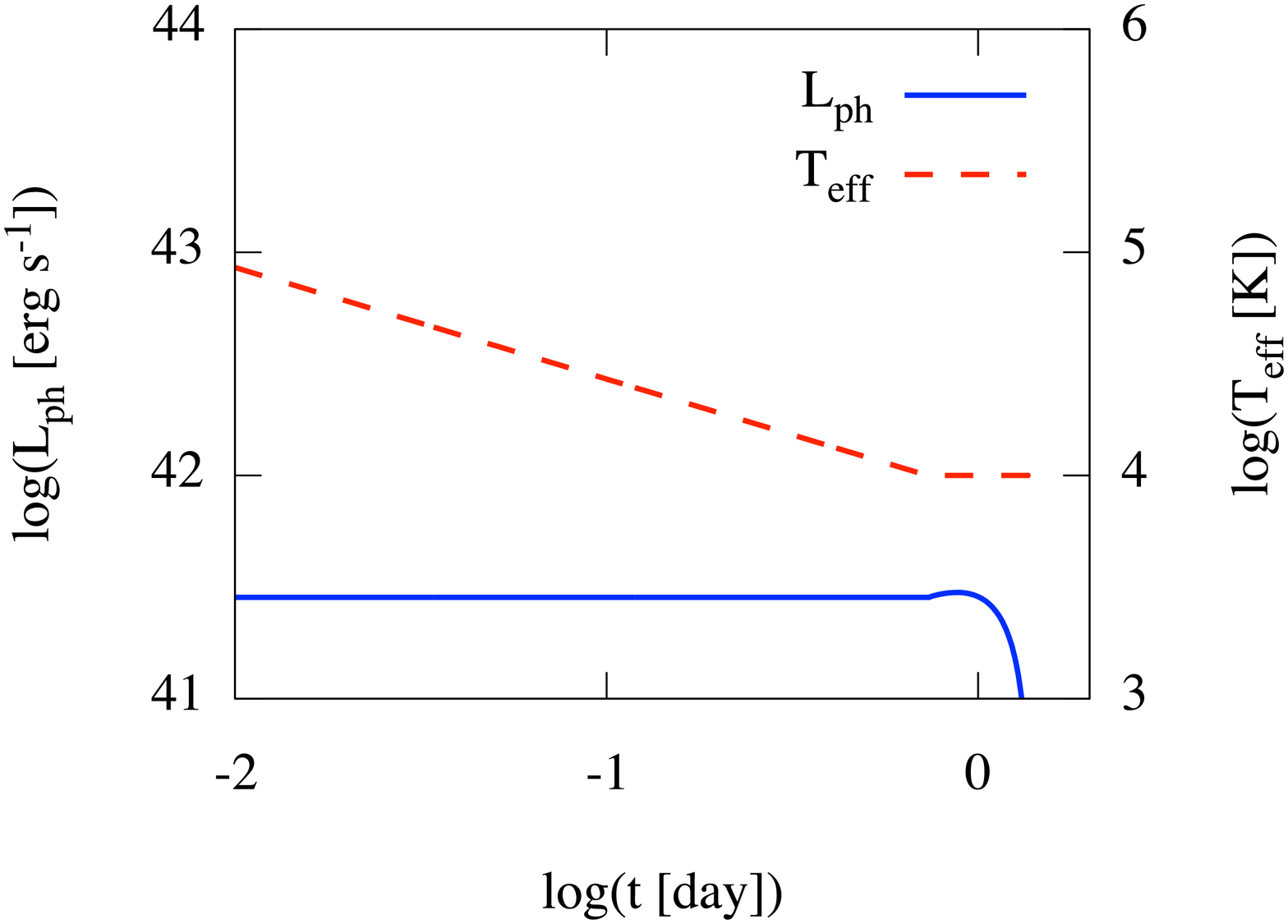}
   \includegraphics[width=\linewidth]{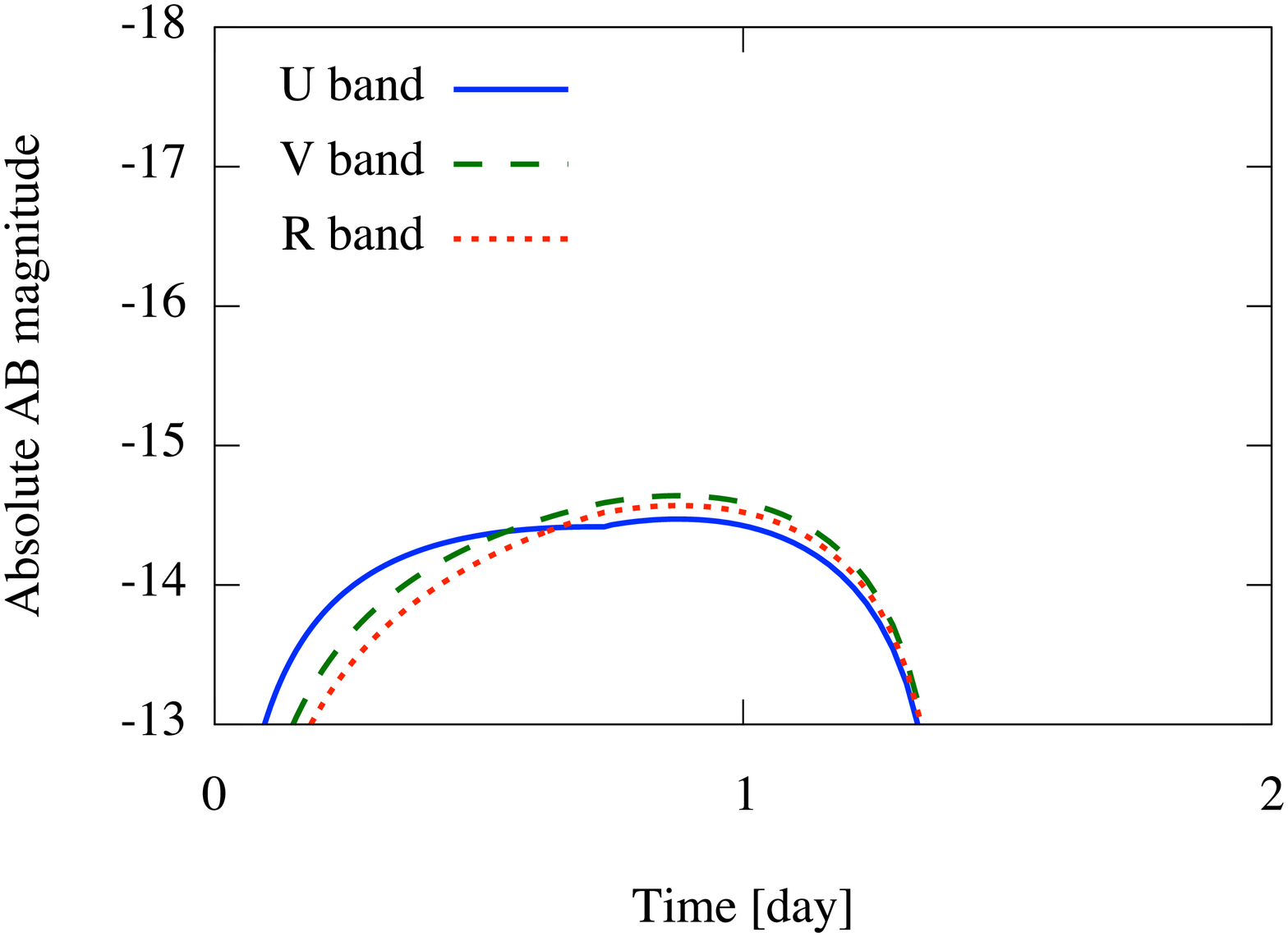}
    \caption{Same as Figure \ref{fig:tidal_evolv}, but for $\eta_w=10^{-1.5}$ and $V_w=10^{9.5}\rm~cm~s^{-1}$.}
    \label{fig:tidal_dim_evolv}
   \end{center}
  \end{figure}

We consider a tidally-synchronized binary system where the spin period of the secondary is synchronized to the orbital period of the primary. 
As binary parameters, we choose the mass of the WR, $M_*=10^{1.5}\rm~\msun$, the radius of the WR, $R_*=10^{11}$ cm, the binary separation, $a=10^{12}$ cm, and the mass ratio, $q=M_*/M_{\rm BH}=1$
\footnote{There is some uncertainty for radii of WR stars. A relation $R_*\sim 7\times10^{10}(M_*/10~\msun)^{0.7}\rm~cm$ is obtained by stellar evolution models~\citep{SM92a,KZK16a}, while $R_*\sim 2\times10^{11}$ cm are proposed from an atmospheric model \citep{Cro07a}.
Besides,  the radius is larger for lighter secondaries, $R_*\sim 10^{12}$ cm for $M_*\sim5\msun$, according to a binary evolution model \citep{YWL10a}.}. 
This parameter set satisfies $t_{\rm TL}\lesssim t_{\rm life}$, and is consistent with stellar evolution models \citep{SM92a}. 
The spin angular velocity is synchronized to the orbital motion, which is estimated to be $\omega_s=\sqrt{2GM_*/a^3}\simeq9.2\times10^{-4}M_{*,1.5}^{1/2}a_{12}^{-3/2}~{\rm s}^{-1}$. 
This value is so high that the outer part of the stellar material cannot fall towards the BH directly. Thus, an accretion disk is formed around the newborn BH. 
This produces a massive outflow, which leads to a TLSSN.

When the secondary collapses, the outer material of the secondary at a cylindrical radius $\varpi$ is at the centrifugal radius, 
$r_{\rm cf}(R)=\varpi^4\omega_s^2/(GM_R)\approx 2\varpi^4/a^3$, 
where $M_R\sim M_*$ is the mass enclosed inside the spherical radius $R=\sqrt{\varpi^2+z^2}$. 
Setting $r_{\rm cf}=6GM_*/c^2$, we obtain the critical radius for disk formation:
\begin{equation}
 R_{\rm cr}\approx\left(\frac{3GM_*a^3}{c^2}\right)^{1/4}\simeq6.1\times10^{10}M_{*,1.5}^{1/4}a_{12}^{3/4}\rm~cm.
\end{equation}
Since $R_*>R_{\rm cr}$ for our parameter choice, 
the outer material can form a disk. The density profile of WRs can be expressed as a polytropic sphere of index $\sim$ 2.5--3.5 \citep{KZK16a}. 
The outer region ($R\gtrsim R_*/2$) of a polytrope $n$ can be fitted as $\rho_{\rm env}\approx\rho_*(R_*/R-1)^n$ \citep{MM99a},
where $\rho_*\approx A_\rho M_*/(4\pi R_*^3)$ and $A_\rho\simeq3.9$ is numerical constant that depends on the polytrope index (we use $n=3$). 
This expression can reproduce the polytrope within errors of a few percents, except for the very outer edge which does not affect the result. 
The disk mass is then estimated to be
\begin{eqnarray}
 M_d&\approx&4\pi\int_{R_{\rm cr}}^{R_*}d\varpi\int_0^{\sqrt{R_*^2-\varpi^2}}dz\varpi\rho_*\left(\sqrt{\frac{R_*^2}{\varpi^2+z^2}}-1\right)^n\nonumber\\
&=&A_\rho I_d M_*\simeq 0.41M_{*,1.5}I_{-2.5}\rm~\msun, 
\end{eqnarray}
where $I_d = \int_{x_{\rm cr}}^1dx\int_0^{\sqrt{1-x^2}}dy x(1/\sqrt{x^2+y^2}-1)^n\simeq3.3\times10^{-3}$ and $x_{\rm cr}=R_{\rm cr}/R_*$. 
Note that $I_d$ is a strong function of $x_{\rm cr}$ that depends on $M_*$, 
so that the dependence of $M_d$ on $M_*$ is not simple. 
The outer material falls to the disk in a free fall time, 
$t_{\rm ff}\approx\sqrt{R_*^3/(GM_*)}\simeq5.4\times10^2M_{*,1.5}^{-1/2}R_{*,11}^{3/2}$ s. 
Then, the mass accretion rate is estimated to be
\begin{equation}
 \mdotenv\approx\frac{M_d}{t_{\rm ff}}\simeq7.5\times10^{-4}M_{*,1.5}^{1/2}R_{*,11}^{-3/2}M_{d,-0.39}\rm~\msun~s^{-1}.\label{eq:mdotenv}
\end{equation}
This accretion rate is much higher than the Eddington rate, $\dot M_{\rm Edd}=L_{\rm Edd}/c^2\simeq2.2\times10^{-15}M_{*,1.5}\rm~M_{\odot}~s^{-1}$, while it is much lower than the critical mass accretion rate for neutrino cooling, $\dot M\sim 1\rm~M_{\odot}~s^{-1}$ \citep{PWF99a,KM02a}.  Then, the physical state of the accretion flow is expected to be the advection dominant regime, where the outflow is likely to be produced \citep{ny94,BB99a,KNP05a}. The wide-angle outflow production is also commonly seen in numerical simulations for central engine of gamma-ray bursts (GRBs) \citep{MW99a,FM13a}.
The outflow luminosity is estimated to be
\begin{eqnarray}
 L_w&\approx&\frac{1}{2}\eta_w\mdotenv V_w^2\\\label{eq:lw_tidal}
  &\simeq&2.4\times10^{49}M_{*,1.5}^{1/2}R_{*,11}^{-3/2}M_{d,-0.39}\eta_{-0.5}V_{10}^2  \rm~erg~s^{-1},\nonumber
\end{eqnarray}
where we assume that the $\eta_w\sim 1/3$ of the accreting material is ejected as the outflow with velocity $V_w\sim 10^{10}\rm~cm~s^{-1}$. Although these parameters related to the outflow are highly uncertain, these values of $\eta_w$ and $V_w$ are consistent with the recent simulation and observation results \citep{HOD15a,TOK16a,NSS17a}.
\footnote{Note that these simulations and observations are for the cases with $\dot M\sim10^2-10^3\dot M_{\rm Edd}$. The values of $\eta_w$ and $V_w$ for $\dot M\sim 10^{10} \dot M_{\rm Edd}$ should be investigated in the future.}
The duration of the outflow is comparable to the free-fall time, 
since the accretion time after the disk formation is much shorter than the free-fall time. 
The total mass of the outflow is $M_w=\eta_wM_d\simeq0.14M_{d,-0.39}\eta_{-0.5}\rm~\msun$ and the total energy is 
\begin{equation}
 {\mathcal E}_w\approx\frac{1}{2}\eta_wM_dV_w^2\simeq1.3\times10^{52}M_{d,-0.39}\eta_{-0.5}V_{10}^2  \rm~erg.
\end{equation}
%Since this outflow is much more massive and energetic than the ejecta of failed SNe, we can neglect the effects of the ejecta for $t\gtrsim10$ s, where $t=0$ is the time when the outflow begin to be produced.

We assume that the outflow occurs at the escape velocity, and is launched at $r_{\rm lp}\approx 2GM_*/V_w^2\simeq8.4\times10^{7}M_{*,1.5}V_{10}^{-2}$ cm.
Assuming that the radiation energy is comparable to the kinetic energy at the launching point, 
the temperature at that point is 
\begin{eqnarray}
 T_{\rm lp}&\approx&\left(\frac{\mdotw V_w}{8\pi a_rr_{\rm lp}^2}\right)^{1/4}\\
  &\simeq&1.4\times10^9M_{*,1.5}^{-3/8}R_{*,11}^{-3/8}M_{d,-0.39}^{1/4}\eta_{-0.5}^{1/4}V_{10}^{5/4}\rm~K,\nonumber
\end{eqnarray}
where $\mdotw=\eta_w\mdotenv$. 
We assume a constant outflow velocity, which leads to $R\sim V_wt$. 
For $t<t_{\rm ff}$, the outflow is continuously produced.  
Considering adiabatic expansion, we obtain 
$\rho_w\propto R^{-2}$ and $T_w\propto R^{-2/3}$ where $\rho_w$ and $T_w$ are the density and temperature in the outflow, respectively. 
At $t\sim t_{\rm ff}$, the accretion rate decreases and the powerful outflow stops.
The radius of the outflow at $t=t_{\rm ff}$ is $R_{w,0}\approx V_wt_{\rm ff}$. The total internal energy at that time is 
\begin{eqnarray}
\mathcal{E}_{\rm int,0}=4\pi\int_{r_{\rm pl}}^{R_{w,0}}a_rT_w^4 r^2 dr\approx \frac{3}{\sqrt 2}\eta_w M_d \frac{GM_*}{R_*}.
\end{eqnarray}
The bulk of photons are trapped inside the outflow, but a small fraction of the photons can escape from the outflow through a diffusion process.
The luminosity of diffusing photons is estimated to be 
\begin{equation}
 L_{\rm ph,0}\approx \frac{\mathcal{E}_{\rm int,0}}{t_{\rm ph,0}} \simeq 9.0\times10^{41} M_{*,1.5}^{1/2}R_{*,11}^{1/2}V_{10}\rm~erg~s^{-1},
\end{equation}
where $t_{\rm ph,0}\approx 9\kappa\eta_wM_d/(4\pi^3 R_{w,0}c)$ is the photon diffusion time at $t=t_{\rm ff}$. 
Since the optical depth is large enough, the diffusing photons have a Planck spectrum with an effective temperature of
\begin{equation}
 T_{\rm eff,0}=\frac{L_{\rm ph,0}}{4\pi \sigma R_{w,0}^2}\simeq 8.5\times10^4M_{*,1.5}^{3/8}R_{*,11}^{-5/8}V_{10}^{3/4}\rm~K,
\end{equation}
where $\sigma$ is the Stefan–Boltzmann constant.

For $t > t_{\rm ff}$, the outflow decouples from the disk and expands in a homologous manner. 
Then, the density of the outflow evolves as $\rho_w\propto t^{-3}$. 
Considering adiabatic expansion, $T_w\propto \rho_w^{1/3}\propto t^{-1}$,
which leads to $\mathcal{E}_{\rm int}\propto R_w^3 T_w^4\propto t^{-1}$.
The diffusion time evolves as $t_{\rm ph}\approx 9\eta_wM_d/(4\pi^3 c R_w)\propto t^{-1}$.
Thus, the luminosity of diffusing photons is constant, $L_{\rm ph}\propto t^0$. 
The effective temperature then evolves $T_{\rm eff}\propto t^{-1/2}$.

When the effective temperature drops to $10^4$K,
the ionized helium ions inside the outflow start to recombine and become neutral \citep[e.g.][]{KK14a}.
The time at which the recombination surface deviates from the outflow surface is estimated to be \citep{Pop93a}
\begin{equation}
 t_i\approx \left(\frac{L_{\rm ph,0}}{4\pi \sigma T_{\rm ion}^4 V_w^2}\right)\simeq 3.6\times10^4M_{*,1.5}^{1/4}R_{*,11}^{1/4}V_{10}^{-3/2}\rm~s,
\end{equation}
The position of the recombination surface is given by \citep{Pop93a}
\begin{equation}
 R_i\approx V_w^2\left[tt_i\left(1+\frac{t_i}{3t_a^2}\right)-\frac{t^4}{3t_a^2}\right],
\end{equation}
where $t_a=\sqrt{2t_{\rm ph,0}R_{w,0}/V_w}$ is the photon breakout time without the recombination surface \citep{Arn80a}.
Here, we assume $t_i\ll t_a$ when estimating $t_i$.
Since the opacity of the neutral gas is very small \citep[e.g.,][]{KK14a}, 
the photosphere is equal to the recombination surface.
Thus, the effective temperature is $T_{\rm eff}=T_{\rm ion}$, 
and the luminosity is estimated to be $L_{\rm ph}=4\pi \sigma T_{\rm ion}^4R_i^2$.
Since $t_i^2 \ll 3t_a^2$ is satisfied for the parameter range of our interest, 
the luminosity has a maximum value of 
\begin{eqnarray}
 L_{\rm pk}&\approx& 4\pi \sigma T_{\rm ion}^4 V_w^2 \left(\frac{3t_it_a^2}{4}\right)\\
&\simeq&5.4\times10^{42}M_{*,1.5}^{1/6}R_{*,11}^{1/6}M_{d,-0.39}^{2/3}\eta_{-0.5}^{2/3}V_{10}^{1/3}\rm~erg~s^{-1}, \nonumber
\end{eqnarray}
and  the time of the peak luminosity is
\begin{eqnarray}
t_{\rm pk}&\approx&\left(\frac{3t_it_a^2}{4}\right)^{1/3}\\
&\simeq& 8.7\times10^4M_{*,1.5}^{1/12}R_{*,11}^{1/12}M_{d,-0.39}^{1/3}\eta_{-0.5}^{1/3}V_{10}^{-5/6}\rm~s. \nonumber
\end{eqnarray}
For $t > t_{\rm pk}$, $R_i$ rapidly decreases, and accordingly the luminosity decays quickly.
The optical depth for the outflow becomes lower than unity in this phase. 
The approximate use of the Planck distribution would become inaccurate, especially at $\tau \lesssim1-10$. 
To study the features of the decay phase, a more careful treatment of thermalization processes would be required.

The evolution of $T_w$ and $L_{\rm ph}$ are shown in the upper panel of Figure \ref{fig:tidal_evolv}, where we use the fiducial parameter set ($M_*=30\rm~M_{\odot}$, $R_*=10^{11}$ cm, $a=10^{12}$ cm, $\eta_w=10^{-0.5}$, $V_w=10^{10}\rm~cm~s^{-1}$).  
The lower panel of the figure shows the evolution of the absolute AB magnitude in the $U$ (365 nm), $V$ (550.5 nm), and $R$ (658.8 nm) bands. 
Since the outflow parameter is uncertain, we show the results for the case with $\eta_w=10^{-1.5}$ and $V_w=10^{9.5}\rm~cm~s^{-1}$ in Figure \ref{fig:tidal_dim_evolv} for comparison.
We can see that the optical band light curves rapidly become bright in several hours, remain bright and slowly varying for days, and then rapidly fade on a timescale of hours.
The peak magnitude range is -14 to -17, which is similar to that of usual type II or type Ib/Ic SNe.
However their shorter durations are useful for distinguishing TLSSNe from the usual SNe. 
Spectroscopic observations can also discriminate TLSSNe from macronovae/kilonovae, since TLSSNe will show strong helium lines while macronovae/kilonovae are not expected to show such lines.
We note that these TLSSNe are bright and short duration, compared to the PIAT events discussed in Paper II. 

Interestingly,  $t_{\rm pk}$ and $L_{\rm pk}$ do not have a strong dependence on  the parameters. However, the occurrence of TLSSNe is sensitive to the value of $x_{\rm cr}=R_{\rm cr}/R_*\simeq0.61M_{*,1.5}^{1/4}R_{*,11}^{-1}a_{12}^{3/4}$. For $x_{\rm cr}>1$, an accretion disk is not formed, which leads to a result similar to the PIATs that we discuss in Paper II. For $x_{\rm cr}\ll 1$, a significant fraction of the stellar material falls onto the disk, and the newborn BH has a large spin, probably resulting in a GRB. 
See Section \ref{sec:summary} for discussion about possible relation between GRBs and TLSSNe.

\section{Radio afterglows of TLSSNe}\label{sec:radio}

Outflow-driven transients may lead to radio afterglows (see \citealt{KHM17a} for details; see also~\citealt{MKM16a}). We briefly discuss this possibility here~\citep[cf.][for supernovae and neutron star mergers]{Che98a,NP11a}.  The deceleration radius and time 
is estimated to be 
\begin{eqnarray}
 R_{\rm dec}&\approx&\left(\frac{3M_w}{4\pi n_{\rm ext}m_p}\right)^{1/3}\\
&\simeq& 5.0\times10^{18}M_{d,-0.39}^{1/3}\eta_{-0.5}^{1/3}n_{-1}^{-1/3}\rm~cm,\nonumber
\end{eqnarray}
\begin{equation}
 t_{\rm dec}\approx\frac{R_{\rm dec}}{V_w}\simeq 5.0\times10^{8} M_{d,-0.39}^{1/3}\eta_{-0.5}^{1/3}V_{10}^{-1}n_{-1}^{-1/3}\rm~s,
\end{equation}
where $n_{\rm ext}=0.1~n_{-1}\rm~cm^{-3}$ is the number density of the circum-binary medium. The deceleration time can be shorter for smaller $\eta_w$ and larger $V_w$. After the deceleration time, the evolution of the outflow is represented by the self-similar solution,
\begin{equation}
R = R_{\rm dec}\left(t/t_{\rm dec}\right)^{2/5}  (t\geq t_{\rm dec})
,\label{eq:sedov-R}
\end{equation}
\begin{equation}
V = 0.4 V_w \left(t/t_{\rm dec}\right)^{-3/5}  (t\geq t_{\rm dec})
.\label{eq:sedov-v}
\end{equation}

We estimate the physical quantities around $t\sim t_{\rm dec}$ using $V\sim V_w$ and $R\sim R_{\rm dec}$. 
The magnetic field is estimated to be $B=(9\pi m_p n_{\rm ext}\epsilon_B v^2)\simeq1.4 V_{10}
n_{-1}^{1/2}\epsilon_{-2}^{1/2}$ mG, where $\epsilon_B$ is the energy fraction of the magnetic field. 
The minimum Lorentz factor of electrons is approximately
\begin{equation}
 \gamma_m\approx \frac{\zeta_e}{2}\left(\frac{m_p}{m_e}\right)\left(\frac{V}{c}\right)^2\simeq41 V_{10}^{2}\zeta_{-0.4},
\end{equation}
where $\zeta_e\sim(p-2)\epsilon_e/((p-1)f_e)\sim 0.4$, $\epsilon_e$ is the energy fraction of the non-thermal electrons, $f_e\sim0.1$ is the number fraction of non-thermal electron, and $p$ is the spectral index of the non-thermal electrons. 
The cooling Lorentz factor is $\gamma_c\approx 6\pi m_e c/(\sigma_TB^2t_{\rm dec})\simeq 7.4\times10^5M_{d,-0.39}^{-1/3}\eta_{-0.5}^{-1/3}V_{10}^{-1}n_{-1}^{-2/3}\epsilon_{-2}^{-1}$, where $\sigma_T$ is the Thomson cross section. 
Since $\gamma_m \ll \gamma_c$, the synchrotron spectrum is in the slow cooling regime.  
The synchrotron frequencies for the electrons of $\gamma_m$ and $\gamma_c$ are 
\begin{equation}
  \nu_m\approx\frac{\gamma_m^2 e B}{2\pi m_e c}
\simeq6.7\times10^6V_{10}^{5}n_{-1}^{1/2}\epsilon_{-2}^{1/2} \zeta_{-0.4}^2\rm~Hz,
\end{equation}
\begin{equation}
 \nu_c\simeq2.2\times10^{15} M_{d,-0.39}^{-2/3}\eta_{-0.5}^{-2/3}V_{10}^{-3}n_{-1}^{-5/6}\epsilon_{-2}^{-3/2}\rm~Hz. 
\end{equation}
If we ignore synchrotron self absorption (SSA), the synchrotron spectrum has a peak at $\nu_m$ and 
its flux is
\begin{eqnarray}
 F_{\nu,m}&\approx& \frac{P_m N_e}{4\pi \nu_m d_L^2}\\
& \simeq&22M_{d,-0.39}\eta_{-0.5}V_{10}n_{-1}^{1/2}\epsilon_{-2}^{1/2}f_{-1}d_{27}^{-2}\rm~mJy,\nonumber
\end{eqnarray}
where $P_m\approx \gamma_m^2\sigma_T c B^2/(6\pi)$ is the synchrotron radiation power per electron, 
$N_e\approx4\pi R^3n_{\rm ext}f_e/3$ is the total non-thermal electron number, and $d_L\simeq10^{27}$ 
cm is the luminosity distance. Since $F_\nu \propto \nu^{(1-p)/2}$ for $\nu_m<\nu<\nu_c$ without SSA, 
the observed flux at frequency $\nu_{\rm obs}$ is estimated to be
\begin{eqnarray}
 F_{\nu,\rm obs}&\approx& F_{\nu,m}\left(\frac{\nu_{\rm obs}}{\nu_m}\right)^{(1-p)/2}\simeq 0.15\nu_9^{1-p\over2}M_{d,-0.39}\nonumber\\
&\times&\eta_{-0.5}V_{10}^{5p-3\over2}n_{-1}^{p+1\over4}\epsilon_{-2}^{p+1\over4}f_{-1}d_{27}^{-2}\rm~mJy,
\end{eqnarray}
where we use $p=3$ to estimate the value. 
Since the sensitivity of current radio surveys is around 0.1 mJy, it is possible to detect this radio emission. 
If the optical transient discussed in Section \ref{sec:optical} is observed, the deeper radio follow-up observation with a sensitivity of around $\rm\mu$Jy can be performed. 
In this case, we can expect detection of the radio signal even with much lower $\eta_w$. 
The deceleration time is shorter for lower $\eta_w$, which helps the coincident detection.
Using Equations (\ref{eq:sedov-R}) and (\ref{eq:sedov-v}), we obtain $F_{\nu,\rm obs}\propto t^{3}$ for $t<t_{\rm dec}$ and $F_{\nu,\rm obs}\propto t^{(21-15p)/10}$ for $t>t_{\rm dec}$. Note that $F_{\nu,\rm obs}$ has a strong dependence on
$V_w$, $F_{\nu,\rm obs}\propto V_w^6$ for $p=3$. Thus, just a few times lower $V_w$ would make it difficult to detect the radio afterglow.

The optical depth for SSA is estimated to be $\tau_a\approx A_p e f_en_{\rm ext} 
R(\nu/\nu_m)^{-(p+4)/2}/(B\gamma_m^5)$, where $A_p$ is a function of $p$ ($A_p=26.31$ for $p=3$, 
see \citealt{MTO14a}). The SSA frequency at which $\tau_a=1$ is estimated as
\begin{eqnarray}
\nu_a &\approx& \left(A_p \frac{ef_en_{\rm ext}R}{B\gamma_m^5}\right)^{2/(p+4)}\nu_m\simeq 7.1\times10^{7} M_{d,-0.39}^{2\over p+4}\nonumber\\
&\times&\eta_{-0.5}^{2\over p+4}V_{10}^{5p-2\over p+4}n_{-1}^{3p+14\over 6p+24}\epsilon_{-2}^{p+2\over 2p+8}\xi_{-0.4}^{2p-2\over p+4}f_{-1}^{2\over p+4}\rm~Hz.
\end{eqnarray}
This frequency evolves as $\nu_a\propto t^{2/(p+4)}$ for $t<t_{\rm dec}$ and $\nu_a\propto 
t^{-(3p+2)/(p+4)}$ for $t>t_{\rm dec}$ \citep{NP11a}. If we focus on $\nu_{\rm obs}>\nu_a$, we can 
ignore the effect of SSA. Since we typically expect $\nu_a > \nu_m$, the spectrum is modified by SSA 
as $F_\nu\propto \nu^{5/2}$ for $\nu_m < \nu<\nu_a$ and $F_\nu\propto \nu^2$ for $\nu<\nu_m$.

\section{Summary \& Discussion}\label{sec:summary}
 
We investigated outflow-driven transients from newborn binary black holes formed from BH-WR binaries, within the context of isolated binary evolution scenarios. 
When the binary separation is small or the binaries are massive enough, the spin period of the WR is synchronized to the orbital period. 
When the WR collapses to a BH, the outer region of the WR has such a high angular momentum that an accretion disk is formed around a newborn secondary BH. 
This results in an energetic outflow of kinetic energy of $\sim 10^{52}$ ergs for $\eta_w\sim10^{-0.5}$, leading to a TLSSN whose bolometric luminosity can be $\sim10^{42}-10^{43}\rm~erg~s^{-1}$. 
Its optical band absolute magnitude reaches $\sim-17$, with a duration of around a day. 
Transient radio emission can also be expected, owing to the large amount of kinetic energy involved.

When the binary separation is larger or the stellar mass is lower, the tidal synchronization may not occur and the spin of the secondary is likely to slow down. Even in this case, a fraction of the outer material of the secondary is ejected when the secondary collapses to a BH. This ejected material is expected to be accreted by the primary BH, and a powerful outflow is produced, leading to a PIAT. We discuss this type of transient in the accompanying paper (Paper II).

The TLSSNe can be distinguished from usual SNe by their shorter duration, 
and from macronovae/kilonovae by their strong helium lines. 
The light curves of TLSSNe are consistent with some of the rapid transients observed \citep{DCS14a,TTM16b} on the basis of their timescale (around a day) and absolute magnitude ($\sim-16$),
although other phenomena, such as shock breakouts from cooling envelopes \citep[e.g.][]{WMC07a} or the outflow-driven transient from single BH formation \citep{KQ15a} could also appear similar.

The current optical surveys with a sensitivity of $\sim 21$ mag, such as Pan-STARRS \citep{Pan-STARRS04a}, PTF  \citep{PTF09a}, and KISS \citep{KISS14a}, imply a detectability distance for TLSSNe of $\sim 200$ Mpc.
Assuming that the event rate of TLSSNe is similar to the merger rate of BBHs, $\sim10-200\rm~Gpc^{-3}~yr^{-1}$ \citep{LIGO16e,LIGO17a}, 
the event rate within the sensitivity range is 0.3--7 yr$^{-1}$.
Thus, the current surveys could detect this type of transients in the near future.
However, we should note that the event rate of TLSSNe has substantial uncertainties, related to the binary evolution and the outflow from a super-Eddington accretion flow.
Future projects, e.g., the Large Synoptic Survey Telescope \citep[LSST,][]{LSST09a}, would be able to detect them or put a meaningful limit on the event rate.

The stellar wind, the ejected stellar envelopes during the common envelope phase, and/or the supernova impostors can significantly pollute the circum-binary medium \citep[e.g.][]{SLS11a,BHB16a}. 
This circum-binary matter can be more massive than the outflow of TLSSNe. 
Thus, they could affect both the optical and radio light curves of TLSSNe. 

Since the core of the WR can be radiative, it is possible that the core rotates faster than the envelope. 
If the core rotates sufficiently fast, it forms a BH with an accretion disk $\sim1$ second after the core-collapse \citep{OO11a}. 
The mass accretion rate is large enough to synthesize some amount of radioactive nuclei around this secondary BH \citep{PWH03a,FHA04a}, which may produce another type of supernova/hypernova-like transient powered by the radioactive decay of nuclei. 
Even relativistic jets that could lead to a GRB may be launched, and during the jet propagation phase, the radioactive nuclei can be synthesized in the shocked envelope, leading to another energy source for supernova-like emission \citep{BDL17a}.
Different energy sources may coexist in the tidally locked system we here consider, so that optical emission from TLSSNe may be powered by either the thermal emission from the disk outflow itself or regenerated emission from radioactive nuclei.

If the core commonly rotates very fast, such a binary system may be responsible for long GRBs \citep[cf.][]{FH05a}. Indeed, the true event rate of GRBs after beaming correction is consistent with the expected event rate of TLSSNe. In addition, the tidally locked system can naturally produce a late-time central engine activity for plateau emission or X-ray flares in GRB afterglows, because the outer envelope accretes onto the BH $\sim10^3$ s after the core accretion that is attributed to the GRB prompt emission.
Note that a disk state for a typical TLSSN discussed in this paper is different from that of a collapsar disk discussed in the context of GRBs \citep{MW99a}. The disk in a TLSSN has much lower temperature than that of a collapsar disk, so a TLSSN disk cannot produce a jet through the neutrino annihilation \citep{ELP89a,PWF99a}. If the secondary BH has a high spin and global magnetic field, the magnetic jet can be produced \citep{BZ77a,Kom04a,TT16a}. Although the jet power seems too low to produce a typical long GRB, it may be observed as an ultra-long GRBs if the jet is directed to the Earth \citep{QK12a,WH12a}. Since a wide-angle outflow can simultaneously be produced \citep{MW99a}, we may also observe a TLSSN.

An accretion disk around a BH in a BBH is left over after the transients considered here.
A few years later, this disk is expected to become a fossil disk, in which the angular momentum transport is inefficient, due to radiative cooling \citep[e.g.,][]{PDC14a,PLG16a,KTT17a}. 
Since such fossil disks can remain for millions of years, a possible outcome from them would be electromagnetic counterparts of the GWs from the eventual BBH mergers \citep{MKM16a,KTT17a,MK17a}.

Besides the transients discussed here, which involve a WR companion,
there are likely to be other formation channels of BBHs through binary evolution, 
where the progenitor consists of a BH and a blue-super giants (BSGs) or red-super giants (RSGs).
Since BSGs and RSGs have larger radii than WRs, $x_{\rm cr}<1$ is easily satisfied. 
In this sense, TLSSNe are likely in BH-BSG and BH-RSG binaries.  
However, whether the spin is tidally synchronized or not depends strongly on the internal structure of the secondary \citep{KZK16a}. 
Also, the tidal force from the primary distorts the WR star to non-spherical shape, which could affect the stellar structure.
A more accurate modeling will require solving the stellar evolution in detail. Due to these
uncertainties related to the stellar structure as well as the outflow properties resulting from the super-Eddington accretion, it is currently difficult to derive a meaningful luminosity distribution for such transients.

\acknowledgments
The authors thank Kazumi Kashiyama, Kunihito Ioka, and Tomoya Kinugawa for useful comments. This work is partially supported by Alfred P. Sloan Foundation (K.M.), NSF Grant No. PHY-1620777 (K.M.), NASA NNX13AH50G (S.S.K. and P.M.), an IGC post-doctoral fellowship program (S.S.K).

%% The reference list follows the main body and any appendices.
%% Use LaTeX's thebibliography environment to mark up your reference list.
%% Note \begin{thebibliography} is followed by an empty set of
%% curly braces.  If you forget this, LaTeX will generate the error
%% "Perhaps a missing \item?".
%%
%% thebibliography produces citations in the text using \bibitem-\cite
%% cross-referencing. Each reference is preceded by a
%% \bibitem command that defines in curly braces the KEY that corresponds
%% to the KEY in the \cite commands (see the first section above).
%% Make sure that you provide a unique KEY for every \bibitem or else the
%% paper will not LaTeX. The square brackets should contain
%% the citation text that LaTeX will insert in
%% place of the \cite commands.

%% We have used macros to produce journal name abbreviations.
%% \aastex provides a number of these for the more frequently-cited journals.
%% See the Author Guide for a list of them.

%% Note that the style of the \bibitem labels (in []) is slightly
%% different from previous examples.  The natbib system solves a host
%% of citation expression problems, but it is necessary to clearly
%% delimit the year from the author name used in the citation.
%% See the natbib documentation for more details and options.
%\bibliographystyle{apj}

%\bibliography{astro}
%\bibliographystyle{apj_8}

%\begin{thebibliography}{}
%\bibliography{ref}

%% Include this line if you are using the \added, \replaced, \deleted
%% commands to see a summary list of all changes at the end of the article.
%\listofchanges

\end{document}